\documentstyle[12pt,epsf]{article}
\begin{document}

\begin{titlepage}
~\vspace{2.5cm}

\vskip2cm
\begin{center}
\Large\bf Accelerated convergence of perturbative QCD by
optimal conformal mapping of the Borel plane
\end{center}

\vspace{1.0cm}

\begin{center}
Irinel Caprini\\
National Institute of Physics and Nuclear Engineering, POB MG 6, Bucharest,\\
R-76900 Romania \end{center}
\centerline{and}

\begin{center}
Jan Fischer\\
Institute of Physics, Academy of Sciences of the Czech Republic, 
182 21  Prague 8, \\ Czech Republic
\end{center}
\vspace{1.2cm}
\begin{abstract}
The technique of conformal mappings is applied to enlarge the  convergence
 domain of the Borel series and to accelerate the convergence of
 Borel-summed Green functions in perturbative QCD.
We use the optimal mapping, which takes into account the location of all the 
 singularities of the Borel transform as well as the
 present knowledge about its behaviour near the first  branch points.
The determination of $\alpha_s(m_\tau^2)$ from the hadronic decay  rate of 
the $\tau$-lepton is discussed as an illustration  of the method.
\end{abstract}
\end{titlepage}
\newpage
\section{Introduction}
The behaviour of the large order terms in perturbative quantum field theory
 has been a subject of permanent interest \cite{Lipa}-\cite{tHof}. Recently, 
this problem received much attention in the case of QCD \cite{Muel}-\cite{BBK}.
As is known, the creation of instanton-antiinstanton pairs and certain classes 
of Feynman diagrams are responsible for a factorial increase of the large 
order coefficients of the  perturbative expansion of the QCD Green functions, 
making this series divergent in the mathematical sense. Moreover, the growth 
of the large-order coefficients is so dramatic that, when combined with other 
difficult circumstances (like their non-alternating sign and the 
extraordinarily small analyticity domain of the Green functions in the 
coupling variable \cite{tHof}), it leads to the situation that some of the 
usually efficient summation  techniques 
 \cite{Hardy} are not applicable. One of them, the Borel summation, was 
 very much investigated in the recent time. The growth of the 
large order perturbative coefficients of the QCD Green functions leads, 
under certain conditions, to Borel transforms with singularities in the 
Borel plane that make the integral defining the Green function by the 
Laplace transform ill-defined. Of course, as discussed
 in \cite{Jan}, the Borel technique is not the only mathematical method by
 which a divergent series can be summed, but an ambiguity emerges in every
 summation method, once it is discovered in one of them.
 The real source of nonuniqueness consists in a missing piece of
 information about the quantities to be calculated, which adopts different
 forms in different summation methods, but has to be added to eliminate 
the ambiguity.

 In QCD the Borel nonsummability
 originates from the infrared regions of the Feynman diagrams,
 where nonperturbative  effects play also an important role.  
Therefore it is natural to assume that the ambiguities of
 perturbation theory  must be compensated by nonperturbative contributions. 
 In fact, it turns out that general concepts like analyticity, 
 renormalization group or specific properties of the QCD vacuum are 
 unavoidable  when discussing the large order behaviour of perturbation 
 theory \cite{Pari},\cite{tHof},\cite{Jan}. As an example, we recall that 
 the argument given by 't Hooft \cite{tHof} for the Borel nonsummability of 
 QCD relies on nonperturbative properties, mainly the momentum plane
 analyticity combined with renormalization group invariance.
 So the perturbative and  the nonperturbative regimes of the theory cannot be
separated, and their interplay is very clear when attempting to
 perform the summation of the large orders of perturbation theory.
The Borel plane is particularly suitable for discussing this aspect, since the
singularities of the Borel transform offer a very intuitive measure of the
 ambiguities of the perturbation theory and suggest the way to compensate
 them. The properties of these singularities  in some approximations (like 
  massless QCD in the large $\beta_0$ limit, when they
 are poles \cite{Bene}) were used  recently to provide estimates of the
 truncation error in the theoretical determinations of some accurately
 measured quantities. 

A natural question is whether it is possible to improve the accuracy 
 of the Borel summation using the first Taylor coefficients of the Borel 
 transform known from the explicit low order calculations, supplemented 
 with some (approximate) information about its singularities in the Borel
 plane. We address this question in the present paper. We use as input the 
 assumption that, for suitable QCD amplitudes, there are no other 
 singularities in the Borel plane than those located on the real axis, at a 
 nonvanishing distance from the origin. The precise nature and strength of 
 these singularities is not known in general, except for the nearest ones, 
 which can be characterized (at least approximately) by using  general 
 principles. As discussed in \cite{Pari},  the singularities of the Borel 
 transform require the introduction of higher dimensional operators, which  
 ensure the compensation of the ambiguities present in the usual perturbative 
 terms by the ambiguities inherent in their Wilson coefficients. This allows 
 one also to infer a universal behaviour of the Borel transform 
near the first ultraviolet renormalon \cite{BBK}. On the other hand, as
 discussed in \cite{Muel}, the location and nature of the first infrared
 renormalon can be plausibly predicted too, by nonperturbative arguments.   
In our approach, we take as input  this  knowledge about 
the first ultraviolet and infrared renormalons.

Our purpose was to exploit in an optimal way this information, 
in order to improve the accuracy of the Borel summation in the frame 
of a specific prescription of handling the singularities of the Borel 
transform. To this end we use the analytic continuation of the Borel 
transform outside the circle of convergence of the Taylor expansion, 
achieved by the technique of conformal mapping. As is known, the conformal 
mappings are very suitable for accelerating the convergence of 
power series. The existence of an optimal
expansion, with the largest convergence domain and  
 the best asymptotic  convergence rate, was proven in  \cite{CiFi} 
 a long time ago. The method is applicable if  
 the position  of the singularities of the function to be approximated 
is known or can be reasonably guessed, which is the case
 in many situations  in particle physics.

  In  the context of the Borel summation in
quantum field theory, the conformal mappings were first considered
 in \cite{Loef}-\cite{Khuri2}. More recently, the method was  
applied in \cite{Alta}-\cite{Soper} to the Borel transform of
 QCD Green functions, following a suggestion made in \cite{Muel}. The 
 purpose was to estimate (and possibly reduce) the
influence of the first ultraviolet renormalon and of the associated 
power corrections on observable quantities,
 by using a variable in which this singularity is pushed
further away from the integration range of the Borel transform. 
However, from the point of view of the convergence rate 
 the mapping used in \cite{Alta}-\cite{Soper} is not optimal, since it
 takes into account only a part of the singularities of the Borel transform,
the ultraviolet (UV) renormalons. By using an optimal treatment, which
 takes into account also the infrared (IR) renormalons and  
 the behaviour near the first singularities, an increased convergence rate 
 and consequently a smaller truncation error are to be expected. 

The objective of our work is to establish whether the optimal mapping 
technique is numerically  relevant in the Borel plane for situations of 
physical interest (we mention alternative attempts to
 enlarge the convergence domain of the Borel transform,
 based on Pad\'e approximants \cite{Ellis}). To 
illustrate our discussion we consider, as in \cite{Alta}, 
 the Adler function of the massless QCD vacuum polarization and the
 determination of the strong coupling constant
 $\alpha_s(m_\tau^2)$ from the hadronic $\tau$ decay rate.
In the next section we briefly review
some properties of the Adler function and of
 its Borel transform. In section 3 we present
 the technique of optimal conformal mapping and investigate
its efficiency in the Borel plane on several mathematical models which
simulate  the physical situation.
We discuss also the determination of the strong coupling
constant $\alpha_s(m_{\tau}^2)$ 
using the present technique.
Some conclusions are formulated in section 4.

\section{ The Adler function and its Borel transform}
We consider the correlator
 \begin{equation}\label{2point}
   i\int{\rm d}^4 x\,e^{i q\cdot x}\,\langle 0|\,T\,\{
   V^\mu(x), V^\nu(0)^\dagger\}\,|0\rangle
   = (q^\mu q^\nu - g^{\mu\nu} q^2)\,\Pi(s)
  \,,  \end{equation}
  where $s=q^2$ and
  $V^\mu=\bar q\gamma^\mu q$ is the current of a massless quark.
 From the general principles of causality and unitarity it follows that the
amplitude $\Pi(s)$ is an analytic function  of real type in
the complex plane $s$, cut along the real positive axis from the threshold
$4m_\pi^2$ of hadron production to infinity.  It is convenient to define the
Adler  function \begin{equation}\label{Ddef}
   D(s) = -{{\rm d}\over{\rm d}s}\,\Pi (s)\,,
\end{equation}
 which is ultraviolet finite and is also analytic in the complex $s$-plane
cut above the unitarity threshold. This function  was much investigated lately 
in connection with  the determination of the strong coupling
constant  $\alpha_s(m_\tau^2)$ from the hadronic decay of
the $\tau$ lepton \cite{Max},\cite{BBB}, \cite{NaPi} -  \cite{Matt1}. The hadronic decay width,
normalized to the leptonic one, is defined as
\begin{equation}\label{Rtau} 
   R_\tau
= {\Gamma(\tau\to\nu_\tau + \mbox{hadrons})\over   
\Gamma(\tau\to\nu_\tau\,e\,\bar\nu_e)}    = \int\limits_0^{m_\tau^2}\!{\rm
d}s\,    {{\rm d}R_\tau(s)\over{\rm d}s} \,,
\end{equation}
where the inclusive hadronic spectrum ${\rm d}R_\tau(s)/{\rm d}s$ is related
to the spectral part of the correlator (\ref{2point}):
\begin{equation}\label{imexp}    {{\rm d}R_\tau(s)\over{\rm d}s}
   = {3 (1+\delta_{\rm EW})\over \pi m_\tau^2}\,\bigg( 1 - {s\over m_{\tau}^2}
   \bigg)^2  \bigg( 1 + {2 s\over m_\tau^2} \bigg)\,
   \mbox{Im}\,\Pi(s+i\epsilon)\,.
\end{equation}
The factor $\delta_{\rm EW}\simeq 0.0194$ accounts for electroweak
radiative corrections. The decay rate (\ref{Rtau}) was measured recently with
great accuracy \cite{Aleph}, \cite{Cleo}.

 Using the analyticity properties of the function $\Pi(s)$ in the 
 momentum plane, the relation  (\ref{Rtau}) can be
transformed by a Cauchy relation into 
\begin{equation}
\label{dcircle}
   R_\tau
   = {3(1+\delta_{\rm EW})\over 2\pi i}\,\oint {{\rm d}s\over s}
\bigg( 1 - {s\over m_\tau^2}
   \bigg)^3  \bigg( 1 + { s\over m_\tau^2} \bigg)\,
   \,D(s)\,,\end{equation}
where the integration runs along a closed contour in the complex plane,
 taken usually to be the circle $\vert s\vert =m^2_\tau$.

The relation (\ref{dcircle}) is the starting point for the computation of the
$\tau$ hadronic width in perturbative QCD. At complex values of $s$ the Adler
function admits the formal renormalization-group-improved 
expansion 
\begin{equation}\label{Dseries}
   D(s) = 1 + \sum_{n=1}^\infty\,D_n\,\bigg( {\alpha_s(-s)\over\pi}
   \bigg)^n \,.
\end{equation}
The strong coupling $\alpha_s(\mu^2)$ satisfies the 
renormalization-group equation
 \begin{equation}\label{alpha}
 \mu^2{{\rm d}\alpha_s (\mu^2)\over
{\rm d}\mu^2}=- \alpha_s (\mu^2)\sum_{n=0}^\infty
\beta_n (\alpha_s (\mu^2))^{n+1}\,, \end{equation}
with the first coefficients $\beta_n$ defined 
in terms of the number $n_f$ of quark flavours as
\begin{eqnarray}\label{betan}
\beta_0&=&{33-2n_f\over 12 \pi}\nonumber\\
\beta_1&=&{153-19n_f\over 24 \pi^2}\,.
\end{eqnarray}
The coefficients  $D_n$ in the expansion (\ref{Dseries}) were
computed for $n\le 3$  \cite{Chet}-\cite{Surg}. In the $\overline{\rm MS}$
scheme with $n_f=3$ they are
\begin{eqnarray}\label{D1D2D3}
   D_1 &=& 1 \,, \nonumber\\
   D_2 &=&  1.63982 \,, \nonumber\\
   D_3^{\overline{\rm MS}} &=&  6.37101 \,.
\end{eqnarray}
On the other hand, the large-order coefficients,
 $D_n$ for large $n$, have the generic factorial behaviour 
\begin{equation}\label{D_n}
D_n\approx C_k\, n !\, n^{\delta_k}\,\left({\pi \beta_0 \over k}\right)^n+....
\end{equation}
where  
 the index $k$ takes the values: $-1, \pm 2, \pm 3....$.

In the Borel method of summation  one defines the Borel transform of
the Adler function as 
\begin{equation}\label{Bseries}
   B(u) = \sum_{n=0}^\infty\,b_n\,u^n
   \,,
\end{equation}
where
\begin{equation}\label{bn}
b_n= {1\over n! }\,{D_{n+1}\over (\pi \beta_0)^{n}}\,=\,{\widetilde
D_{n+1}\over n!} \,.
\end{equation}
Then $D(s)$ can be expressed formally in terms of $B(u)$
 by the Laplace transform 
\begin{equation}\label{Laplace}
   D(s) = 1 + {1\over\pi \beta_0}\,
    \int\limits_0^\infty\!{\rm d}u\, B(u)\,\exp
    \bigg(-{ u\over \beta_0 \alpha_s(-s)}\bigg) \,.
\end{equation}
To illustrate our technique we shall use also a Borel-summed 
expression for the hadronic decay rate
$R_\tau$. Such an expression was obtained in \cite{Alta} 
by inserting in (\ref{dcircle}) the
Borel representation (\ref{Laplace}) of  the Adler function and performing the
integration along the circle $|s|=m_\tau^2$ 
in the one-loop approximation
\begin{equation}\label{adef}
  \alpha_s (-s)\,={1 \over \beta_0 \ln(-s/ \Lambda^2)}\,
\end{equation}
of the running coupling.
 This procedure gives \cite{Alta}
\begin{equation}\label{Rtaub}
R_\tau\,=\,3 (1+\delta_{{\rm EW}})\left[1+{1\over \pi \beta_0}\int_0
^\infty{\rm d}u \exp\bigg(-{u\over \beta_0\alpha_s(m_\tau^2)}\bigg)\,B(u)\,
F(u)\,\right]\,, \end{equation}
where
\begin{equation}\label{F}
F(u)={-12\,\sin(\pi u)\over \pi\,u(u-1)(u-3)(u-4)}\,.
\end{equation}
We shall use (\ref{Rtaub})  as a starting point for a
 determination of $\alpha_s(m_\tau^2)$ in Section 3.
 
The growth (\ref{D_n}) of the Taylor coefficients $D_n$
leads to the dominant behaviour 
\begin{equation}\label{Bpoles}
B(u)\approx C_k\, \Gamma\,(\delta_k+1)\,(1-{u\over k})^{-\delta_k-1}+... , 
\end{equation}
which shows that the function $B(u)$ becomes singular at the points 
$u=k$, with $ k=-1, \pm 2,
\pm 3...$. The precise values of  $C_k$ and $\delta_k$ are  not known in
general. However, from general arguments it was shown that the nature of the
first branch   points of  the Borel transform  is universal 
\cite{Muel}, \cite{BBK}. More precisely,
near  the first UV renormalon at $u=-1$ the Borel transform
behaves as 
\begin{equation}\label{uv}
B(u)\,\simeq\, r_{1}\,(1+u)^{-\gamma_1}\,,
\end{equation}
where \cite{BBK}
\begin{equation}\label{gama1}
\gamma_1=3-{\beta_1\over \beta_0^2}+\lambda_1\,.
\end{equation}
Here $\lambda_1$ is a parameter depending on the number of flavours, which 
reflects the mixing of higher dimensional operators in the renormalization
 group equations \cite{BBK}.
 Similarly,  near the first IR renormalon at $u=2$ the behaviour is
\begin{equation}\label{ir}
B(u)\,\simeq \,r_2\, (2-u)^{-\gamma_2}\,\,,
\end{equation}
where \cite{Muel} 
\begin{equation}\label{gama2}
\gamma_2=1+2{\beta_1\over \beta_0^2}\,.
\end{equation}
Using the first coefficients $\beta_i$ from (\ref{betan}) 
  and the parameter $\lambda_1$ given in \cite{BBK} (equal to 0.379
for  $n_f=3$ and 0.630 for $n_f=5$) we obtain
\begin{equation}\label{gamanum3}
\gamma_1=2.589\,,\,\,\gamma_2=2.580
\end{equation}
for $ n_f=3$, and
\begin{equation}\label{gamanum5}
\gamma_1=2.972\,,\,\, \gamma_2=2.316
\end{equation}
for $n_f=5$.
We emphasize that only the nature of the first renormalons is known, and
nothing can be said about the residues $r_1$ and $r_2$  appearing in 
(\ref{uv}) and {\ref{ir}, respectively.

Strictly speaking, the integrals (\ref{Laplace}) or
(\ref{Rtaub}) have nothing to do with the  summation of the perturbative
series, because one condition of the Borel   theorem (the existence of the
analytic continuation in the $\alpha_s$ plane from the convergence disk to
an infinite strip   around the positive real semiaxis) is not satisfied
\cite{Hardy}, \cite{tHof}.   This can be seen from the singularities of the 
 Borel transform
given in (\ref{Bpoles}):  the poles situated on the real positive axis (IR
renormalons) make the   integrals (\ref{Laplace}) and (\ref{Rtaub}) 
ambiguous. In order to
compute them a   prescription has to be adopted, by suitably choosing the
integration  contour in order to avoid the singularities. But this is not the
 Borel summation. Different prescriptions give different results, and a measure
 of the intrinsic ambiguity of the perturbation expansion is given by the
 difference between these results, if no a priori arguments in favor of a
 certain choice exist. 

A prescription adopted by several authors \cite{Khuri2}, \cite{Max}, \cite{BBB}
 is the ``principal value'' (PV), defined as
\begin{equation}\label{PV}
{\rm PV} \int_0^\infty {\rm d}u \exp\left(-{u\over a_s}\right) f(u)\equiv
 {1\over 2}\,\left[\int_{{\cal C_+}}{\rm d}u \exp\left(-{u\over a_s}\right)
 f(u)\,
{\rm d}u \,
+ \int_{{\cal C}_-}{\rm d}u \exp\left(-{u\over a_s}\right)f(u)\right] \,,
\end{equation}
where ${\cal C}_\pm$ are two lines parallel to the real axis, slightly above
and below it.  This definition is a generalization to arbitrary singularities
of the Cauchy principal value of simple poles, and has the
advantage of yielding a real result when $a_s$ is real. Although this
prescription is not unique, we  adopted it as a working hypothesis. 

We applied the definition (\ref{PV}) for computing the Borel-summed  
Adler function (given by the Laplace integral (\ref{Laplace})),
and  the hadronic $\tau$ decay rate (given by (\ref{Rtaub})).
In the first case the parameter $a_s$ is related to the running coupling
$\alpha_s(-s)$ (and may be complex if $s$ is complex or in the Minkowskian
region $s>0$), while in the second case it is proportional to 
$\alpha_s(m_\tau^2)$. We use as input the first Taylor coefficients of $B(u)$  
(known  from (\ref{bn}) and the calculated values (\ref{D1D2D3})), 
supplemented by the knowledge  on  the location and the  nature of 
singularities of this function. As discussed in the Introduction, our purpose 
is to improve the accuracy of the calculation by the technique of conformal 
mappings, which exploits in an optimal way this input information.

\section{Optimal conformal mapping of the Borel plane}

\subsection{Remarks on the theory}
The use of conformal mappings for improving the convergence of power
series in particle physics was discussed for the first time
 in Ref. \cite{CiFi}. 
 The problem  formulated in \cite{CiFi} was to find the optimal conformal 
 transformation which minimizes the asymptotic truncation error of a power 
 series, taking into account the location of the singularities of the 
function to be approximated. First we briefly describe the results obtained 
in \cite{CiFi}. We consider a function $f(u)$ analytic in a domain $\cal D$ 
of the complex $u$ plane containing the origin, and write its  Taylor series 
truncated at a finite order $N$ as
\begin{equation}\label{utaylor}
f^{(N)}(u)=\sum_{n=0}^N f_n\,u^n\,.
\end{equation}
According to general theory, the series (\ref{utaylor}) converges, for 
$N\to\infty$, inside the circle passing through the nearest singularity of 
the function $f(u)$ in the complex plane, the rate of convergence at a 
point $u$ situated inside the circle being that of the geometrical series in 
powers of ${r\over R}$, where $r=\vert u\vert$ and $R$ is the radius of
 the convergence circle. Therefore, the convergence rate
 is strongly influenced by the distance of the singularities
of $f(u)$  from the origin, 
and can be improved by using a suitable change of variable, in which the 
singularities are pushed further away from the region of interest.

 Let us consider a conformal mapping $w=w(u)$ of the plane
$u$ onto the plane $w$, such that $w(0)=0$, and
write the truncated Taylor expansion of the function $f(u)$
in the variable $w$:
\begin{equation}\label{wtaylor}
f^{(N)}_{w}(u)=\sum_{n=0}^N c_n\,w^n\,.
\end{equation}
As pointed out in \cite{CiFi}, the best asymptotic rate of convergence
 of this series in a certain region of the complex plane is achieved 
when $w$ is such a transformation of the $u$ plane that the corresponding
 ratio ${r\over R}$ is minimal for every point in that region.
According to the theorem proven in \cite{CiFi} this is realized by
a conformal mapping $w(u)$  which maps the whole analyticity domain 
$\cal D$ of the function $f(u)$ in the $u$ plane into 
the interior of a circle in the plane $w$. The proof of the theorem is based
on the fact that the circle is the domain of convergence for 
power series, and  on the  Schwartz lemma, which implies that
the larger the domain mapped inside the  circle, the better is 
the asymptotic rate of convergence (for details see \cite{CiFi}).

In what follows we shall apply this technique to the Borel transform 
$B(u)$ of the Adler function. The nearest singularities
of the function $B(u)$  are situated at $u=-1$ and $u=2$
and the power expansion (\ref{Bseries}) converges only inside the
circle $\vert u\vert <1$ passing through the first UV renormalon. 
It is easy to see that the optimal 
conformal mapping in the sense explained above is given in our case by
\begin{equation}\label{wborel}
w={\sqrt{1+u}-\sqrt{1-u/2}\over \sqrt{1+u}+\sqrt{1-u/2}}\,.
\end{equation}
By this mapping, the complex  $u$ plane cut along the real axis for $u>2$ and 
$u<-1$ is  mapped onto the interior
of the circle $\vert w\vert\, <\, 1$ in the complex $w$-plane,
 the origin $u=0$ of the $u$ plane
becoming the origin $w=0$ of the $w$ plane,  and
the upper (lower) lips of the cuts are mapped onto the upper
(lower) semicircle in the plane $w$.
Particularly, all the singularities of the Borel transform, the  UV and IR
 renormalons,  are now situated on the boundary of the unit disc in the $w$ 
 plane, all at equal distance from the origin. The Taylor expansion of 
the Borel transform  in powers of  $w$, 
\begin{equation}\label{Bwtaylor}
\tilde{B}_{w}^{(N)}(u)=\sum_{n=0}^N \tilde{c}_n\,w^n\,,
\end{equation}
will converge for $N\to\infty$ up to points close to the renormalons.
This is a considerable improvement with respect to the usual expansion
(\ref{Bseries}), whose convergence in the Borel plane is limited by the 
circle reaching    
the first UV renormalon. The requirement of convergence of (\ref{Bwtaylor}) 
on this disc implies the holomorphy of the expanded function on the disc. 
In this way, the expansion in powers of $w(u)$ makes full use of the 
analyticity property that is universally
 (but tacitly) assumed in all QCD considerations, namely
 that there are no singularities in the Borel plane  other than those situated 
on the real axis, at a nonvanishing distance from the origin. 
 This essential, additional     
 assumption  has, to our knowledge, not been explicitly used. 

For comparison we give the conformal mapping used in \cite{Alta}-
\cite{Soper},
\begin{equation}\label{zborel}
z={\sqrt{1+u}-1\over \sqrt{1+u}+1}\,,
\end{equation}
which maps the $u$ plane cut along $u<-1$ onto the interior of the unit circle
in  the $z$ plane. In the $z$ plane the UV renormalons are situated 
along the boundary of the unit circle $|z|=1$, but the IR renormalons are
situated  inside this circle. As noticed in \cite{Soper}, 
pushing away the ultraviolet renormalons by (\ref{zborel})
 has a price in moving the first infrared renormalon 
(and actually, the whole positive real semiaxis) closer to 
the origin. This is why the convergence domain of the power series in $z$ 
is limited by the  first IR renormalon and, as a consequence, the 
convergence rate of the series in powers $z$ will be worse than that 
obtained with the optimal variable (\ref{wborel}). The use of the optimal 
conformal mapping (\ref{wborel}) is therefore highly desirable, because 
it does not suffer from this shortcoming, placing {\it all} the renormalons 
onto the circumference of the unit disk. 

As was noticed in \cite{CCF}, a further improvement of the convergence 
rate can be reached if some information about the nature of the 
singularities of the expanded function $f(u)$ is available. The idea is that 
the power variable $w(u)$, taken as a function of $u$, should resemble $f(u)$ 
as much as our knowledge of $f(u)$ allows it. (As it was put in \cite{CiFi}, 
if we were to know $f$ exactly, the most rapidly convergent expansion would 
be that in powers of $f$ itself, in which case it would reduce to the 
identity $f \equiv f$.)  

In practice, however, our knowledge of the expanded function is only 
approximative. For instance, as discussed above, we know that
 near the branch points $u=-1$ and $u=2$ the function
$B(u)$ behaves like $(u+1)^{-\gamma_1}$ and  $(2-u)^{-\gamma_2}$ 
respectively, with the $\gamma_i$ real 
positive numbers. In this case it is convenient to expand the product
$(u+1)^{\gamma_1}\,(2-u)^{\gamma_2} \,B(u)$ in powers of the optimal
variable $w$ defined in (\ref{wborel}), and  introduce then explicitly the 
singular factors. The expansion of the function $B(u)$ will have the form
\begin{equation}\label{Bw1taylor}
\hat{B}_{w}^{(N)}(u)={1\over 
 (u+1)^{\gamma_1} (2-u)^{\gamma_2}} \sum_{n=0}^N \hat{c}_n\,w^n\,.
\label{extr}
\end{equation}
 The singularities themselves may survive as positive powers 
$(u+1)^{\gamma_1}$ and  $(2-u)^{\gamma_2}$, the bonus nevertheless 
being that the positive exponents $\gamma_i$ keep the values of the 
function  
\begin{equation}\label{product}
 (u+1)^{\gamma_1} (2-u)^{\gamma_2} B(u)
\end{equation}
finite near $u=u_{1}$ and $u=u_{2}$, 
which softens their numerical impact. 
This step will imply no large-order improvement of the convergence rate
 (because the rate is given by the {\it position} of the nearest
 singularities), unless some of the two singularities is fully
 removed by it. But it may represent a considerable improvement at low
 orders, even if the nature of the nearest singularities 
is known only approximately. A nice example
of efficiency of this approach in practice  was presented by Soper and
Surguladze in \cite{Soper}.   

\subsection{Discussion of mathematical models}
We tested the practical efficiency of the conformal mapping 
(\ref{wborel})  for a number of functions 
having logarithmic or power branch points at $u=-1$ and $u=2$.
We took  functions close to the physical situation as described in 
Section 2, {\it i.e.} 
we started from a "perturbative" expansion of the form
\begin{equation}\label{Buseries}
   B^{(N)}(u) = \sum_{n=0}^N\,b_n\,u^n
   \,,
\end{equation}
with low values of $N$. The expansion (\ref{Bwtaylor}) in terms of the
variable $w$ is  obtained by replacing $u$ in (\ref{Buseries}) with 
the expansion
\begin{equation}\label{cw}
u^{(N)}_w=\sum_{n=1}^N C_n w^n 
\end{equation}  
which follows from the inverse of (\ref{wborel}), and  keeping only terms up to
the order $N$ ({\em i.e.} $(u^{(N)}_w)^N=C_1^N w^N$, etc). The numerical
values of the first   coefficients $C_n$ are:
\begin{equation}\label{cwnum}
C_1={8\over 3}\,,\, C_2={16\over 9}\,,\, 
C_3=-{40\over 27}\,,\, C_4=-{224\over 81}\,,\, C_5=-{88\over 243}\,...
\end{equation}
For comparison, the expansion in powers of
 the variable $z$  given in (\ref{zborel}) is obtained using
\begin{equation}\label{cz}
u^{(N)}_z=\sum_{n=1}^N \bar C_n z^n\,
\end{equation}  
with the numerical values \cite{Alta}
\begin{equation}\label{cznum}
\bar C_1=4\,,\, \bar C_2=8\,,\, 
\bar C_3=12\,,\, \bar C_4=16\,,\, \bar C_5=20\,...
\end{equation}

 We first computed the model functions and their various approximants at 
 points $u$ inside the analyticity region, near the origin of the Borel 
 plane. In most of the  cases investigated the expansions in powers of the 
 optimal variable $w$ approximated the exact functions much better than 
 the standard expansion (\ref{Buseries}) or the series in powers of 
the $z$ variable. This feature was visible even with a few terms 
in the expansion. Actually, as we discussed in the previous 
Section, for the physical applications we are interested in 
the calculation of Laplace integrals like (\ref{Laplace}) or
(\ref{Rtaub}).  We evaluated this integral, with the generalized principal
value prescription  defined in (\ref{PV})
for a large number of functions of physical interest.
 We consider as an example the function 
 \begin{equation}\label{Bmodel}
B(u)={r_1\over (1+u)^{\gamma_1}}+
{r_2\over (2-u)^{\gamma_2}}+\sum_{n=3}^{N_{IR}}{r_n\over (n-u)^{\gamma_n}}\,,
\end{equation}
which simulates the contribution of a few  renormalons.
 The principal value (\ref{PV}) of the Laplace integral was computed
 numerically with great accuracy. In order to check the computations we used
 the relation \cite{Bate}
\begin{equation}\label{ingam}
\lim_{\epsilon\to 0} \int_0^\infty {\rm
d}u {\exp(-u/a_s)\over (u+b+i\epsilon)^\nu}\, =\,a_s^{1-\nu}{\rm e}^{b/a_s}
\Gamma(-\nu+1, b/a_s)\,,\,{\rm Re}\, a_s>0\,, \end{equation}
where $\epsilon>0$  and 
$\Gamma(\nu, z)$ is the incomplete gamma function \cite{Bate},
 analytically continued
 from the region ${\rm Re}\,z\,>0$ to the whole complex plane $z$ cut along the
 negative  real axis. 
For integer $\nu$ this can be expressed equivalently as \cite{Abra}
\begin{equation}\label{expin}
\lim_{\epsilon \to 0}\int_0^\infty{\rm
d}u {\exp(-u /a_s)\over (u+b+i\epsilon)^n}\,
=\,{{\rm e}^{b/a_s} \over  b^{n-1}}E_n(b/ a_s)\,,\,{\rm Re}\,
a_s>0\,, \end{equation}
in terms of the exponential integral functions  $E_n(z)$.
Actually, 
as seen from (\ref{Bmodel}), in the physical case 
the denominators must be defined so as to 
give the correct cut structure of the Borel transform. This case 
 is obtained from (\ref{expin}) as  
\begin{equation}\label{conv}
\lim_{\epsilon \to 0}\int_0^\infty{\rm
d}u\, {\exp(-u/a_s)\over (|b|-u-i\epsilon)^n}\,=\,{{\rm e}^{b/a_s}\over  -
|b|^{n-1}}E_n(b/a_s)\,, 
\end{equation}
where $b=-|b|$.

As a side remark, we mention that the above relations are useful for defining
the principal value prescription for arbitrary values of $a_s$.
First,  by means of  repeated integration by parts in (\ref{expin}) we can
express  the left hand side as \cite{Bate} 
 \begin{eqnarray}\label{expin1}
\int_0^\infty{\rm
d}u {\exp(-u/a_s)\over (u+b+i\epsilon)^n}\,
=\sum_{m=1}^{n-1}{(m-1)!\over (n-1)!}\,{(-a_s)^{1-n+m}\over b^m}\,+
\,{(-a_s)^{1-n} \over (n-1)! }\int_0^\infty{\rm
d}u {\exp(-u/ a_s)\over
(u+b+i\epsilon)}\,,\nonumber \\
{\rm Re}\, a_s>0\,. \end{eqnarray}  
We now apply the definition (\ref{PV}) of the principal value and use 
the symbolic relation
\begin{equation}\label{pmieps}
{1\over (u+b\pm i\epsilon)}={\rm PV}{1\over (u+b)}\,\mp\,i\pi \delta (u+b)\,.
\end{equation}
in the last term in (\ref{expin1}). 
 We obtain thus the following expression of the principal value 
\begin{equation}\label{pvbate}
{\rm PV} \int_0^\infty{\rm
d}u {\exp(-u/ a_s)\over (u+b)^n}\,
=\,{{\rm e}^{b/a_s} \over  b^{n-1}}E_n(b/a_s)\,+\,i\pi{(-a_s)^{1-n}\over
(n-1)!} {\rm e}^{b/a_s}\,,\,\,{\rm Re}\, a_s>0\,.
\end{equation}  
For real values of $a_s$ the last term in the above relation is purely
imaginary and compensates the imaginary part of the first term. In this 
case the definition (\ref{pvbate}) amounts therefore to taking the real 
part of the right hand side of (\ref{expin}). For complex  $a_s$, when 
the last term in (\ref{pvbate}) has also a nonvanishing real 
part, the compensation does not occur, and the result is complex. 
As discussed below Eq.(\ref{PV}),
complex values of $a_s$ appear in the Borel summation of the Green functions
in the complex momentum plane or in the timelike region.  For some Minkowskian
 quantities a definition of the principal value,  
based on physical arguments, was proposed  in  \cite{Max}.  The above
expression  (\ref{pvbate}) is general and covers all these cases.

\begin{figure}
\centerline{ \epsfxsize=8cm\
 \epsffile{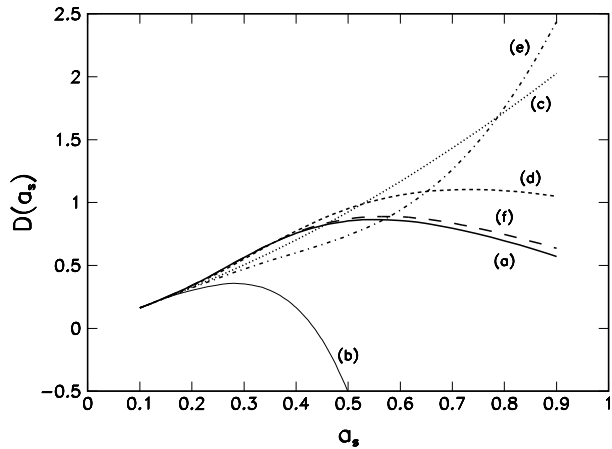}}
\caption{ Laplace integral for the model (\ref{Bmodel}) 
as a function of $a_s$:  exact values (a); perturbation expansion (b);
 expansion in powers of $z$ (c); improved expansion
in powers of $z$ (d); expansion in powers of $w$ (e);
 improved expansion in powers of $w$ (f). The series are truncated at $N=3$.}
\vspace{0.5 cm}
\end{figure}

In Fig.1  we give for illustration the results of our analysis
for the model function (\ref{Bmodel}) with the parameters: $r_1=1, r_2=4,
r_n=0\,,n\ge 3, \gamma_1=2.5,  \gamma_2=2.5$, and $a_s$  in the range
(0.1-0.9). The Laplace integral of  the exact function (\ref{Bmodel}) is
indicated together with the results  given by its ordinary perturbative
expansion (truncated at $N=3$), and the  expansions accelerated by the
conformal mappings $z$ and $w$, both in the  simple versions and with the
improvement explained in (\ref{Bw1taylor}). To simulate the physical situation
in a more realistic way, we assumed that  the nature of the first
singularities is not exactly known   and used in the improved version
(\ref{Bw1taylor}) the product of $B(u)$   with the factors
$(u+1)^{2.0} (2-u)^{2.0}$, which do not compensate exactly the 
singularities of the model function (\ref{Bmodel}). 

As seen from  Fig.1, the combined technique of conformal mapping and the 
explicit treatment of the branch points, supposing that some (approximate) 
information about the behaviour of the function near the first singularities 
is available, improves 		
the accuracy of the Borel integral, 
especially for large $a_s$. The values $a_s\approx 0.2-0.3$ (of interest in 
the hadronic $\tau$ decay) are on the boundary of the region for which the 
improvement is significant at this order, $N=3$. We  notice that
a major part of the improvement is brought by the separate treatment of 
the branch points, according to
 (\ref{Bw1taylor}), especially at low $N$. 
Even the standard expansion in powers of the
Borel variable $u$ gives good results if the nature of
 the lowest singularities, assumed to be exactly known,
is treated explicitly as above. However,  when the behaviour near the first 
 singularities is known only approximately, the expansion in the optimal 
 variable gives in general the best approximation, especially when the 
order $N$ of the truncation is increased. 
\begin{table}
\begin{center}
\begin{tabular}{|c|c|c|c|c|c|c|} \hline 
&&&&&&\\
$N$&$(a)$&$(b)$&$(c)$&$(d)$&$(e)$&$(f)$ \\ \hline
&&&&&&\\
3&0.35&0.530&0.50&0.546&0.47&0.5596\\ \hline
&&&&&&\\
4&0.90&0.610&0.540&0.5625&0.64&0.5701 \\ \hline 
&&&&&&\\
5&--0.10&0.5732&0.547&0.5722&0.5457&0.5701 \\ \hline
&&&&&&\\
6&2.2&0.516&0.5631&0.5783&0.613&0.56518 \\ \hline
&&&&&&\\
7&$-$3.6&0.82&0.5743&0.582&0.63&0.56518 \\ \hline
&&&&&&\\
8&13&$-$0.12&0.587&0.584&0.515&0.563102 \\ \hline
&&&&&&\\
9&$-$40&2.3&0.599&0.584&0.64&0.563100 \\ \hline
&&&&&&\\
10&143&$-$3.4&0.613&0.583&0.503&0.563470 \\ \hline
&&&&&&\\
11&$-$541&9.5&0.63&0.581&0.540&0.563467 \\ \hline
&&&&&&\\
12&2 $10^3$&$-$19&0.64&0.5776&0.590&0.563783\\ \hline
&&&&&&\\
15&$-$2 $10^5$&199&0.72&0.56400&0.5783&0.563713\\ \hline
&&&&&&\\
20&2 $10^9$&$-$8 $10^3$&1.1&0.530&0.582&0.563689\\ \hline
&&&&&&\\
25&$-$4 $10^{13}$&3 $10^5$&2.7&0.48&0.5687&0.563681\\ \hline
&&&&&&\\
30&2 $10^{18}$&$-$5 $10^7$&13&0.43&0.559754&0.563682\\ \hline
\end{tabular}
\caption{ Approximations of  the Laplace integral for $a_s=0.3$ for different
  truncation orders: $(a)$ expansion in powers of $u$; $(b)$ ``improved''
  expansion in powers of $u$; $(c)$ expansion in powers of $z$;
 $(d)$ ``improved'' expansion in powers of $z$; $(e)$ 
expansion in powers of $w$; $(f)$ ``improved''
  expansion in powers of $w$. The exact value is $D(0.3)=0.563683$. 
  The results close to the exact value are indicated
  with a greater number of digits.}
\end{center}  
\end{table}
We illustrate this fact in 
Table 1, where we indicate  the Laplace integral for $a_s=0.3$, of 
the function (\ref{Bmodel}) with
 $\gamma_1=\gamma_2=2.5$  
 as a function of the truncation order $N$, for different types of expansions.
The ``improved'' expansions were obtained now by expanding in powers the
product of $B(u)$   with the factors
$(u+1)^{1.5} (2-u)^{1.5}$, close but
 not identical with the actual behaviour of 
(\ref{Bmodel}). 

For larger values of $a_s$ the improved accuracy obtained by using 
 the optimal mapping is even more impressive.  
Some results are presented in  Table 2 and in Table 3,
  for $a_s=0.5$ and $a_s=0.8$, respectively. Similar results were obtained
 also for model functions with more singularities on the real axis.
\begin{table}
\begin{center}
\begin{tabular}{|c|c|c|c|c|c|c|} \hline 
&&&&&&\\
$N$&$(a)$&$(b)$&$(c)$&$(d)$&$(e)$&$(f)$ \\ \hline
&&&&&&\\
3&$-$0.51&1.04&0.925&1.01&0.739&0.97\\ \hline
&&&&&&\\
5&$-$15&1.9&1.22&1.05&1.7&0.8769 \\ \hline 
&&&&&&\\
10&4 $10^4$&$-$75&2.4&0.8412&2.2&0.857092 \\ \hline
&&&&&&\\
12&2 $10^6$&349&3.6&0.678&1.8&0.853068 \\ \hline
&&&&&&\\
15&$-$9 $10^8$&3 $10^3$&8.2&0.37&1.6&0.852614 \\ \hline
&&&&&&\\
20&8 $10^{13}$&$-$1 $10^5$&56&$-$0.27&2 $10^{-3}$&0.853263 \\ \hline
&&&&&&\\
25&$-$2 $10^{19}$&1 $10^8$&597&$-$1.06&0.59&0.853463 \\ \hline
&&&&&&\\
30&1 $10^{25}$&$-$3 $10^{13}$&8 $10^3$&$-$2.0&1.5&0.853438 \\ \hline
\end{tabular}
\caption{ The same as in Table 1, for $a_s=0.5$. The exact value
  is $D(0.5)=0.853427$.}  
\end{center}
\end{table}

A closer look at the Tables 1, 2, and 3 reveals that there are essentially
three circumstances affecting the convergence properties: (i) the use of a 
convenient (including the optimal) conformal mapping, (ii) explicit (but, in
practice, approximative) account of the branch point singularities, 
and (iii) exponential damping of the integrand by $\exp\left(-{u/a_s}\right)$. 

\begin{table}
\begin{center}
\begin{tabular}{|c|c|c|c|c|c|c|} \hline 
&&&&&&\\
$N$&$(a)$&$(b)$&$(c)$&$(d)$&$(e)$&$(f)$ \\ \hline
&&&&&&\\
3&$-$9.1&1.9&1.7&1.34&1.8&1.00\\ \hline
&&&&&&\\
5&$-$296&6.7&3.2&1.13&6.1&0.61 \\ \hline 
&&&&&&\\
10&8 $10^6$&$-$381&17&$-$0.442&13&0.69528 \\ \hline
&&&&&&\\
12&9 $10^8$&$-$2 $10^3$&41&$-$1.35&$-$18&0.6873 \\ \hline
&&&&&&\\
15&$-$2 $10^{12}$&$$2 $10^4$&178&$-$3.0&$-$22&0.70106 \\ \hline
&&&&&&\\
20&1 $10^{19}$&5 $10^6$&3 $10^3$&$-$6.2&$-$7.1&0.69741 \\ \hline
&&&&&&\\
25&$-$2 $10^{24}$&3 $10^{12}$&6 $10^4$&$-$10&21&0.69580 \\ \hline
&&&&&&\\
30&4 $10^{30}$&$-$3 $10^{18}$&2 $10^6$&$-$14&$-$12&0.696204 \\ \hline
\end{tabular}
\caption{ The same as in Table 1, for $a_s=0.8$. The exact value
  is $D(0.8)=0.696408$.}  
\end{center}
\end{table} 
The effect of the factor (i) can be seen from the fact that, in each of the 
Tables, the column (c) possesses better convergence properties than the 
column (a), and the column (e) has better properties than the column (c). 
The effect of (ii) is seen from the fact that, again in all three Tables, 
the columns (f), (d) and (b) have better convergence properties than the 
columns (e), (c) and (a), respectively. As concerns the point (iii), we see 
from Table 3 that the salutary effect of the optimal conformal mapping 
is most spectacular when the damping of the exponential function 
$\exp\left(-{u/a_s}\right)$ is the weakest, i.e., 
when $a_{s}$ has the highest value, $a_{s}=0.8$ in our case. Indeed, in this
case, only the combined technique of the optimal conformal mapping and the
explicit treatment of the branch points leads to numerical convergence, with
steadily increasing accuracy up to $N=30$ by the least.   

In the case of a stronger damping (Table 1), the role of the optimal conformal 
mapping combined with a careful regard to the branch point singularities is 
again important, but good results are obtained also by the other methods, the 
success varying with the perturbation order $N$ used; see the different 
columns (b) -- (f) of Table 1 at different values of $N$. The asymptotic
superiority of the optimal mapping (columns (e) and (f)) emerges at very high
values of $N$; this mapping supersedes the other methods and 
turns out to be the best at least from $N=20$ on up to the highest value of 
$N$ shown in the Tables, $N=30$. 

It is not excluded, on the other hand, that even the best series, column (f),  
will exhibit numerical indications of divergence at still higher 
orders; note that the singularities survive in some form because we 
(on purpose, in order to simulate real situations) had not completely 
removed them (see(ii)). Consequently, as the Borel integral path runs 
along the cut, which in the $w$ plane is mapped onto the {\it boundary 
circle} of the convergence disk, no convergence is warranted even in the 
column (f). It was already pointed out that for 
lower values of $a_{s}$, where the influence of infrared renormalons is 
more suppressed (see e.g. Table 1), results close to $D(0.3)$ are obtained 
even when other methods are used, see   
Table 1, sometimes only at lower values of $N$. 
   
To further illustrate the use of the optimal conformal mapping introduced 
above, we investigated it in a  model proposed in \cite{Alta}, 
 adjusted to better simulate the physical situation.
We assume the case when
 the  Borel function is exactly given by
\begin{equation}\label{Btrue}
B_{true}(u)=1+\widetilde D_2 u+{\widetilde D_3
\over 2}u^2 +[\widehat B(u)-
\sum_{n=0}^2\hat b_n u^n]\,.
\end{equation}
In this expression  the parameters $\widetilde D_2$ and $\widetilde D_3$
are for the moment arbitrary and
the numbers $\hat b_n$
are the Taylor coefficients of
the expansion of
 $\widehat B$ around the origin:
\begin{equation}\label{Bhatser}
\widehat B(u)=\sum_{n=0}^\infty \hat b_n u^n\,.
\end{equation}
For $\widehat B(u)$  we choose the expression
\begin{equation}\label{Bhat}
\widehat B(u)={B_0(2)\over (1-u/2)}\,+\sum_{l=1}^{N_{UV}}{A_0(l)+A_1(l)u\over
(1+u/l)^2}\,+\,\sum_{l=3}^{N_{IR}}{B_0(l)+B_1(l)u\over (1-u/l)^2}\,,
\end{equation}
with 
\begin{eqnarray}\label{Al}
A_0(l)&=&{8\over 3}{(-1)^{l+1}(3l^2+6l+2)\over l^2(l+1)^(l+2)^2}\nonumber\\
A_1(l)&=&{16\over 3}{(-1)^{l+1}(l+3/2)\over l^2(l+1)^(l+2)^2}\nonumber\\
B_0(2)&=&0\nonumber\\
B_0(l)&=&-A_0(-l)\,\,; ~~~~l\ge 3\nonumber\\
B_1(l)&=&-A_1(-l)\,\,; ~~~~l\ge 3\,.
\end{eqnarray}
The function $\widehat B(u)$ coincides actually with the Borel transform 
in the large $\beta_0$ limit \cite{Max},
with finite numbers
$N_{UV}$ and $N_{IR}$ of UV renormalons  and IR renormalons, respectively.
The meaning of the the model adopted above
for $B_{true}$ is clear: it represents a function with the first 3 Taylor terms
specified explicitly on the right hand side of (\ref{Btrue}), and the higher
order terms arising from the sum of $N_{UV}$ UV renormalons and $N_{IR}$
 IR renormalons.
The perturbative expression of this model is therefore
\begin{equation}\label{Bpert}
B_{pert}(u)=1+\tilde D_2 u+{\tilde D_3\over 2}u^2 \,.
\end{equation}
The expansion to the same order in terms of the optimal
conformal variable $w$ can be obtained easily 
 using (\ref{cw}):
\begin{equation}\label{Bmapw}
B_{pert, w}(u)=1+\tilde D_2\,C_1w+(\tilde D_2\,C_2+
{\tilde D_3\over 2}C_1^2)w^2 \,.
\end{equation}
We consider also, for comparison, the expansion
in terms of the variable (\ref{zborel}) used in \cite{Alta}:
\begin{equation}\label{Bmapz}
B_{pert,z}(u)=1+\tilde D_2\bar C_1z+(\tilde D_2\bar C_2+{\tilde D_3\over 2}
\bar C_1^2)z^2 \,,
\end{equation}
with $\bar C_N$ defined in (\ref{cz}). 
We introduce now the expressions 
$B_{true}\,,B_{pert}\,,B_{pert,w}$
and $B_{pert,z}$ in the Laplace integral (\ref{Laplace}) 
and define the corresponding quantities
$D_{true}\,,D_{pert}\,,D_{pert,w}$
and $D_{pert,z}$. 
The integrals defining $D_{pert}$
and $D_{pert,z}$ are well defined, and for
$D_{true}$ and $D_{pert,w}$ we adopt
the princupal value prescription (\ref{PV}).

Folowing \cite{Alta} we consider the ratio
\begin{equation}\label{H}
H_w\,=\,{D_{true}-D_{pert,w} \over
D_{true}-D_{pert}}\,,
\end{equation}
and the similar quantity $H_z$. Clearly, the inequalities
 $\vert H_w \vert\,<\,1$ (or $\vert H_z \vert\,<\,1$)
are the conditions for the accelerated methods based on the conformal mappings
$w$ (or $z$) to be successful. As in \cite{Alta} we look for the domain
in the plane $\tilde D_2\,,\,\tilde D_3$ for which the accelerated methods
give better results than the usual perturbation theory.
In Figs.2 (a) and (b) we represent these domains, for the conformal mappings
$z$ and $w$, respectively (we used in this example the value $a_s=0.27$).

\begin{figure}[htb]
\centerline{\epsfxsize=7cm\epsffile{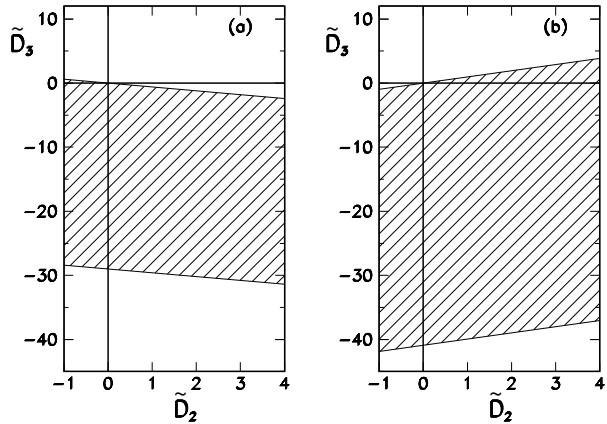}}
\caption{ Bands corresponding to $\vert H\vert <1$ for the conformal mapping 
 $z$ of \cite{Alta} (a) and the optimal conformal mapping
 $w$ defined in the present paper (b).}
\vspace{0.5 cm}
\end{figure}

As seen from Fig.2, the  domains 
 are bands bounded by parallel straight lines,
one of the lines passing through the origin.
The slope  is given by
\begin{equation}\label{slope}
s_w=-{\int_0^\infty {\rm d}u \,{\rm e}^{-u/a_s}[u-C_1 w-C_2 w^2] \over
\int_0^\infty {\rm d}u \,{\rm e}^{-u/a_s}[u^2-C_1^2 w^2]} \,,
\end{equation}
for the conformal mapping (\ref{wborel}) with the coefficients $C_n$ from
(\ref{cwnum}). A similar relation defines the slope $s_z$ in the
case of  the conformal mapping (\ref{zborel}).
As seen from (\ref{slope}) the slopes depend
uniquely by the conformal mapping and not on the details of
the model function. The numerical values obtained from 
(\ref{slope}) are $s_z=-0.60$ and $s_w=0.98$.

As concerns the intercept of the second line defining the allowed domain,
it is given by
\begin{equation}\label{interc}
I_w=
-2{\int_0^\infty {\rm d}u \,{\rm e}^{-u/a_s}[\widehat B(u)-\sum_0
^2\hat b_n u^n]\over
\int_0^\infty {\rm d}u \,{\rm e}^{-u/a_s}[u^2-C_1^2 w^2]} \,,
\end{equation}
and depends on the model function $\widehat B(u)$.
With our choice (\ref{Bhat}) the intercept was rather stable when increasing 
the number of terms $N_{UV}$ and $N_{IR}$ in the expansion. The results presented
in Fig.2 correspond to $N_{UV}=4$ and $N_{IR}=6$.

Fig.2(a) shows, as already remarked in \cite{Alta}, that the conformal mapping
(\ref{zborel}) brings no improvement when the low order coefficients
$\widetilde D_2$ and $\widetilde D_3$ are both positive, as is the case 
of physical interest (see (\ref{D1D2D3})). On the contrary, as shown in Fig.2
(b), there are pairs of positive
($\widetilde D_2, \widetilde D_3$) for which 
 the optimal conformal mapping 
improves the ordinary perturbation expansion
(the point of coordinates $\widetilde D_2=0.724$, $\widetilde D_3=1.23$
obtained from (\ref{Dseries}) and (\ref{Bseries}) is actually
 close to the upper boundary of the domain in Fig.1(b)). Therefore, even at
 very low orders ($N=3$) an improvement can be obtained in principle 
by using the optimal variable. As the first coefficients have no alternate
signs, this might mean that
 the first IR renormalon competes with the first UV one in contributing to
 these coefficients.   
 
 We recall that in the last model we compared only the conformal mappings, 
 without additional information about the nature of the first singularities 
 of the Borel transform.                    

\subsection{Determination of  $\alpha_s(m_{\tau}^2)$ from $\tau$ decay}
As a final application of the method  we discuss the
determination  of the strong coupling constant $\alpha_s(m_{\tau}^2)$ from the
hadronic $\tau$ decay width. It is known that the
theoretical error is at present the dominant ambiguity in this
determination, and the main
source of this error arises from higher orders
in perturbation theory. This makes the
hadronic $\tau$ decay a very suitable place to apply the
 technique of conformal mapping, which accelerates
the convergence of the perturbative expansion and reduces the truncation
error. As we mentioned, this problem was studied previously in \cite{Alta}, where the conformal  
mappings were used to reduce only the effect of the UV renormalons. It is of
interest to use also the optimal conformal mapping, whose properties
were demonstrated on mathematical models. We do not attempt to make here a
complete analysis of $\alpha_s(m_{\tau}^2)$ determination, but only point out
the effect of the combined technique of optimal conformal mapping and the
dimplementation of the correct behaviour of the Borel transform near the first
singularities. 

We  used as starting point the
Borel sum (\ref{Rtaub})  of $R_\tau$  and evaluated 
this  expression using both the
standard Taylor expansion   (\ref{Bseries}) of the Borel transform in powers of $u$, 
and the optimized expression (\ref{Bw1taylor}) proposed by us. For comparison with
previous work, we notice that the "standard expansion"  in our approach 
is equivalent to the 
method of integration along the circle proposed in \cite{LDPi}, 
in the particular case of the one loop running coupling.
The expansions were truncated at $N=2$, with the coefficients $b_n$
determined from (\ref{D1D2D3}) and (\ref{bn}). In the improved
expansion  (\ref{Bw1taylor}) we used the values $\gamma_1$ and $\gamma_2$
given in (\ref{gamanum3}) and the coefficients $\hat c_n$  were computed such
as to reproduce the first three coefficients $b_n$ from (\ref{bn}).

In Fig. 3 we give the results corresponding to the standard
  Taylor expansion (\ref{Bseries}) of the Borel transform 
(curve $(a)$) and the improved expansion
 (\ref{Bw1taylor}) (curve $(b$)),
for various values of $\alpha_s(m_\tau^2)$. 
\begin{figure}[htb]
\centerline{ \epsfxsize=7cm\epsffile{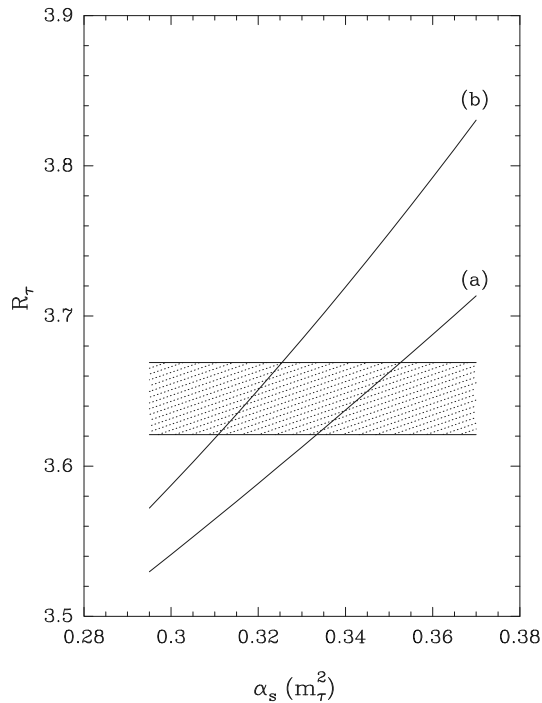}}
\caption{The Borel summation (\ref{Rtaub}) of $R_\tau$, using the standard
  Taylor expansion $ (a)$ and the
 improved expansion (\ref{Bw1taylor}) $(b)$,
as functions of $\alpha_s(m_\tau^2)$.
The band indicates the experimental values. }
\vspace{0.5cm}
\end{figure}
Using the experimental value \cite{Aleph}
\begin{equation}\label{Rtauexp}
(R_\tau)_{exp}=3.645\pm 0.024\,,
\end{equation}
 we obtain
\begin{equation}\label{result}
\alpha_s(m_\tau^2)=0.343\pm (0.009)_{exp}\,,
\end{equation}
using the standard Taylor expansion (\ref{Bseries}), and
\begin{equation}\label{result1}
\alpha_s(m_\tau^2)=0.318\pm (0.007)_{exp}\,,
\end{equation}
using the optimised expression  (\ref{Bw1taylor}). 
 The improved expansion leads to a
value of the coupling constant $\alpha_s(m_\tau^2)$ lower by about 8\%
than the result given by the standard Taylor expansion of the Borel transform. Actually, as in the above
 discussion of the model functions at low $N$ and similar values of $\alpha_s$,
the major contribution in shifting the value of $\alpha_s$ towards smaller
 values is brought by the explicit
 treatment of the first singularities of the Borel transform. At the small
 values of $\alpha_s$ relevant for the present problem, the effect of the 
conformal mapping is barely seen. 

In  (\ref{result}) and (\ref{result1}) we indicated only
 the experimental error, which is very small. On the other hand,
  it is not easy to ascribe a definite
 theoretical error to these results. The problem of the theoretical error
 of $\alpha_s(m_\tau^2)$
  was discussed in many papers,
in particular in  \cite{BBB}, \cite{Alta}, \cite{LDPi}-\cite{Max1}, with
 different conclusions about its magnitude. One can safely neglect
  the effect of the uncertainties in the QCD
 parameters (quark masses, gluon condensate etc), which is small \cite{Braa},
 \cite{Matt1} (leaving aside the still open problem of the $1/s$ corrections). The
 ambiguities related to the prescription chosen for computing the Laplace 
 integral are believed to be small
 too, due to the conjecture that these ambiguities  must be compensated 
 by corresponding ambiguities in
 the condensates. The most important sources of theoretical 
 error remain therefore those related  to 
 the analytic continuation from the euclidian to the minkowskian region,
  and  the truncation of the perturbative expansion.
A complete discussion of these errors is outside the objective 
 of this paper. Concerning the analytic continuation, we only mention that in 
 the derivation of (\ref{Rtaub}) the
perturbative expansion of the Adler function was assumed to be equally valid 
in the euclidian region and in
the complex plane near the timelike axis, which is certainly not true. It is 
not trivial to relax this assumption and see its impact on the determination 
of $\alpha_s(m_\tau^2)$.
  As concerns the truncation error, the
estimate $\delta\alpha_s(m_\tau^2)\simeq 0.05$ was suggested in \cite{Alta}
and \cite{Matt1}, by comparing  the predictions of different
summation procedures. In \cite{Max1} it was claimed on the other hand that
 much less errors are obtained  if the renormalization group invariance
of the perturbation series is exploited in an optimal way. The present
 work points towards a similar conclusion: indeed, as  was remarked also in 
 \cite{BBB}, it is rather arbitrary to interpret the spread of the results 
 produced by different conformal mappings as a measure of the theoretical 
 error, as suggested in \cite{Alta}. Our investigation on mathematical 
 models  shows that the truncation error depends on the choice of the 
 conformal mapping, being smaller if more information on the
analyticity of the function is taken into account. The expansion
proposed in our work exploits in an optimal  way the (renormalization group 
invariant) information on the first renormalons of the Borel transform, and 
we therefore expect that the truncation error of the result (\ref{result1}) 
is smaller than the estimate given above. 

\section{Conclusions}
The technique of the optimal conformal mapping of the Borel plane, 
discussed in this paper, can be seen as an alternative resummation 
of higher-order effects in perturbative QCD. This resummation method has a 
physical content in the sense that the requirement of convergence in powers 
 of the optimal variable $w(u)$ amounts to a statement on {\it analyticity in
the whole double-cut Borel plane}. Indeed, the theorem \cite{CiFi} on the 
asymptotic rate of convergence of power series, on which it is based, is 
dependent upon the condition that the function $f(u)$ (which is expanded) and 
the function $w(u)$ (in powers of which $f(u)$ is expanded, see 
(\ref{wtaylor})) should have the same location of singularities. The method of 
the optimal conformal mapping allows us to make full use of this analyticity
property. 

This remarkable feature is lost if the function is expanded in powers of some
other variable, be it $u$ or a conformal mapping of $u$ such that only a part
of the analyticity domain is mapped inside the convergence circle. In such
cases, the convergence domain is smaller than the region of analyticity, and 
the requirement of convergence has to be supplemented with the analyticity
condition. Only in the case of the optimal mapping the two regions are 
identical. 

As renormalons express the properties of the Feynman diagrams of the process, 
a statement about their location implies a statement about the physics of the 
process considered. 

If the power expansion is truncated at a definite order, as is the 
case in practice, the roles of $u$ and $w(u)$ are modified. While a
polynomial in $u$ is holomorphic in $u$ and has no singularites in the Borel
plane, a polynomial in $w$ has the same analyticity region as the expanded
function, having the cuts equally located. Since singularities have physical 
interpretation, every polynomial in $w(u)$ carries this piece of information.  

We demonstrated the practical use of the optimized expansion numerically 
on a large 
number of model functions, for a sufficiently large truncation 
order ($N\geq 5$). The accuracy of the Laplace integral is especially 
increased by the optimal variable if the coupling constant is large 
and the exponential damping of the integrand is weak. In these cases 
the knowledge (even approximate)  of the behaviour near the first 
renormalons, combined with the expansion in the optimal variable, leads 
to very accurate results, while the expansion in the Borel variable, 
though partially improved by the treatment of the branch points, fails 
dramatically. On the other hand, at low orders of perturbation expansion 
and for values of the coupling constant of physical interest the effect 
of the optimal conformal mapping is not very visible and the predominant 
effect is given by the explicit treatment of the nearest branch points. 
This was actually the case with the determination of the strong coupling 
constant $\alpha_s(m_{\tau}^2)$  from the hadronic $\tau$ decay width: 
the combined technique of conformal mapping and the explicit treatment of 
the first branch points of the Borel transform reduce by about 8\% the 
value given by the usual Taylor expansion in the Borel variable. The major 
contribution to this result is brought by  the theoretical information 
\cite{Muel}, \cite{BBK} about the nature of the first renormalons. 
\vskip0.5cm

{\bf Acknowledgements:}  We are grateful to  Prof. A. de
R\'{u}jula   and the CERN Theory Division for hospitality while a part of
this work was done. One of us (I.C.) thanks Prof. H. Leutwyler for his kind
hospitality at the Institute of Theoretical Physics,
 University of Berne, and the Swiss National
Science Foundation  for support in the program CSR CEEC/ NIS, Contract No 7
IP 051219. The other author (J.F.) is indebted to Prof. G. Altarelli for
reading the manuscript and stimulating discussions and remarks. The work was 
supported in part by GAAV and GACR (Czech Republic)
under grant numbers  A1010711 and 202/96/1616 respectively.


\end{document}